\documentclass[namedreferences]{kluwer}
\usepackage[]{graphicx}
\begin{document}
\begin{article}

\begin{opening}
\title{Pulsating sdB Stars: A New Approach to Probing their Interiors}

\author{Steven D. \surname{Kawaler}}
\author{Shelbi R. \surname{Hostler}}
\runningauthor{S. \surname{Kawaler} \& S. \surname{Hostler}}
\runningtitle{Interiors of sdB stars}

\institute{Dept. of Physics and Astronomy, Iowa State University,
Ames, IA  50011 USA} 

\begin{abstract}

Horizontal branch stars should show significant differential rotation
with depth.   Models that assume systematic angular momentum exchange in the
convective envelope and local conservation of angular momentum in the core
produce HB models that preserve a rapidly rotating core.  A direct probe
of core rotation is available.  The nonradial pulsations of the EC14026
stars frequently show rich pulsation spectra.  Thus their pulsations probe
the {\it internal} rotation of these stars, and should show the effects
of rapid rotation in their cores.  Using models of sdB stars that include
angular momentum evolution, we explore this possibility and show that some
of the sdB pulsators may indeed have rapidly rotating cores.

\end{abstract}

\end{opening}

\section{Introduction}

First studied in detail by R. Peterson in the mid 1980s
(Peterson 1983, 1985a,b), the rapid rotation of some
horizontal branch stars remains an enigmatic feature of this phase of
evolution.  Observations by Behr et al. (2000a,b) 
show a
gap in the horizontal branch of M15 and M13, with two distinct rotation
rates on either side.  Cooler than about 11000~K,
HB stars frequently show rapid rotation ($v \sin i$ $\approx $ 40 km/s)
while no rotation rates above 10 km/s appear in the hotter stars.

One class of field stars is clearly related to horizontal branch stars.
The subdwarf B (sdB) stars are very hot ($T_{\rm eff}\gtrsim 25,000K$),
and show high gravity ($\log g \approx 5.3 - 6.1$). 
Models proposed to explain their origin include common envelope binary
evolution (eg. \opencite{sandtaam}, \opencite{mengetal76}), 
variations of mass loss
efficiency on the RGB \cite{dcruz}, and even mass stripping by planetary
companions \cite{SokHar00}.   Some proposed origin scenarios
make clear predictions about their rotational properties; for example,
those that involve binary mergers or common envelope evolution should spin
up the sdB product.  Extrapolation of the results for Pop. II stars
suggests that sdB stars should be slow rotators, with $v \sin i \lesssim$
10km/s.  The work by \inlinecite{hebetal99} has shown
that at least one star, PG~1605+072, is a more rapid rotator -- as expected
based on an asteroseismic analysis of the pulsations by \inlinecite{Kaw98}.


\inlinecite{SP2000}
(hereafter SP2K) explored HB rotation and single-star evolution
with different assumptions about internal angular momentum transport.
Starting with models that are rotating as solid bodies at the departure from
the main sequence, SP2K followed evolution of the models through the helium
flash and onto the HB.  They studied several limiting cases of
angular momentum transport, two of which we follow up on in our study of
Pop. I hot HB stars.  Both conserve specific angular momentum
(angular momentum per unit mass, $j$) in convectively stable regions.
In convective regions, convective mixing is assumed to produce either
constant angular velocity in the convection zone (Case B) or constant $j$
(Case D); in both cases, the angular momentum contained within the convection
zone is preserved.

SP2K concluded that the best fit to observations of
Pop. II HB rotation came from precursors with rapidly rotating
cores and constant specific
angular momentum in convection zones (i.e. Case D).  These models
could rotate at the rates seen in the cool HB stars.  The slow rotation
in the hotter HB stars requires choking of angular momentum transport by
chemical composition gradients produced by diffusion - such effects are seen
in HB stars (and models) at temperatures above 11,000K (SP2K).
All rotating models of HB stars show 
rapidly rotating cores.  Similarly, sdB stars that originate through binary
mergers or common envelope evolution should show pathological rotation.

A subset of sdB stars are multiperiodic pulsating stars.
The pulsating sdBV stars, if they represent ``typical'' sdB stars and,
by extension, typical HB stars, should allow asteroseismology to address
the state of internal rotation of HB stars.
In fact, one sdBV star, PG~1605+072, was found to have rapid rotation based on the
asteroseismic analysis by \inlinecite{Kaw98} -- this was later partially
verified by \inlinecite{hebetal99} through measurement of rotational
broadening of spectral lines.  Based on this initial success, it is
likely that asteroseismology can be a useful tool to explore important
questions of the second parameter problem in globular clusters and the
origins of field sdB stars.


The  sdB stars probably originate
from higher mass progenitors than Pop  II HB stars.  Stars that are now sdB stars
could have had much higher initial angular momenta \cite{Kaw87}.
Therefore, we have computed models of rotating RGB and sdB stars using Population
I progenitors which
sample the various structures that might be relevant for the pulsators.
We look at the asteroseismic consequences of these rotation profiles
on the observed sdB stars pulsations.

\section{Evolutionary Model Calculations}

The evolution of our stellar models was computed using ISUEVO,
a standard stellar evolution code that is optimized
for producing models for use in asteroseismology studies.  ISUEVO computes
models ranging from the ZAMS through the RGB and AGB and beyond that have been used
for studies of HB stars \cite{Stoetal97, vanhetal98, Kaw98, Kilk1605, ReedetalF48}.

As we are modeling Population I stars, we assume a metallicity Z=0.02
and an initial mixture that is solar.  Model sequences were computed with
initial masses of 1.0, 1.5, and 1.8 $M_{\odot}$, and evolved from the ZAMS
until the helium core flash.
The core masses at the helium flash for these models was 0.4832$M_{\odot}$,
0.4798$M_{\odot}$, and 0.4652$M_{\odot}$ respectively (measured to the
midpoint of the hydrogen--burning shell).  We follow the conventional
technique of computing a ZAHB model \cite{Swegro76}.  For
envelopes thinner than 0.0003$M_{\odot}$, we had to
reduce the thickness of the hydrogen to helium transition zone to preserve
the surface abundance of hydrogen as the pre-flash value.

The transport of angular momentum within stars that are evolving up
the giant branch is dominated by the deep
convective envelope.  
The distribution of specific angular momentum $j$
is affected only by slow diffusive processes
in the absence of convection
\cite{Zahn92}.  Within convection zones, we follow the procedure
of SP2K and examine the two limiting cases of $j$ transport described above. 
The angular velocity profiles in our models represent the extreme
of no diffusion of $j$; in real stars, the angular velocity
profiles will be similar to or shallower than what we report.  The total
angular momentum within radiative zones is also an upper limit, as diffusive
transport of $j$ usually drains angular momentum from the core into the more
slowly spinning envelope.  We note that we calculate the angular momentum profile as a separate side calculation, using the local moment of inertia of each mass shell.  Since all angular velocities are much smaller than breakup velocities, the structural effects of rotation are second--order perturbations and can safely be ignored in these pilot calculations. 

We assume that the initial
rotation on the main sequence was as a solid body.  The initial rotation
rates for our models are taken from \inlinecite{Kaw87} -- for example, for
the 1.8$M_{\odot}$ model, \inlinecite{Kaw87} implies $v_{\rm rot}=185$~km/s,
$J_{\rm init}=3.2 \times 10^{50}$, and $\Omega=2.4 \times 10^{-5}s^{-1}$.
Given that our plan is to use these
models for asteroseismology, the structure of the internal rotation profile
is the focus of our study rather than the predicted surface rotation velocities.


During early RGB evolution, the convective envelope deepens.
A steep angular velocity gradient develops quickly as the star
reaches the RGB, steepening significantly in a few times $10^7$ years.
Evolution up the RGB means an increase in radius (and moment of
inertia) meaning that the outer layers slow down.  At the same time, the
core is contracting and therefore spinning up.  Thus evolution up the RGB
steepens the angular velocity profile.  
On the RGB the dominant feature of the internal rotation
profile is the change in angular velocity that occurs near $0.27M_{\odot}$
from the center of the model.  The rotation profile drops precipitously
beyond as the result of draining of angular momentum from the core
by the convective envelope.

Horizontal branch stars evolved from Case B RGB models show much smaller 
surface rotation velocities than those with Case D precursors.  All models
preserve a rapidly rotating core with roughly constant angular velocity out
to about 0.3$M_{\odot}$.
The rotation profiles of models descended from Case D RGB evolution
contain relics of the high specific angular momentum of material in the
outer layers of the RGB progenitor.
The angular velocity profile
changes little during EHB evolution - the core does spin down a bit as the
angular momentum is mixed, but the inner core continues to spin about a
factor of 5 slower than the fastest material.  The angular velocity drops
from the maximum by a factor of about 10 or so to the surface.
\begin{figure}
\centerline{\includegraphics[width=4in]{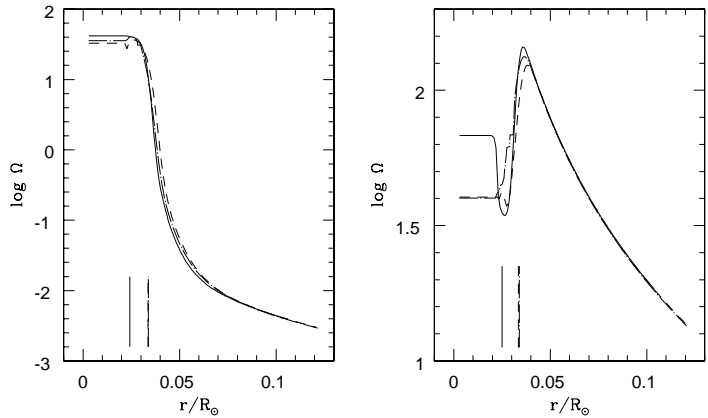}}
\caption{ Rotation profiles for EHB models.  
Left: Case B evolution. Right: Case D evolution.
Models at three stages of evolution are shown; vertical
lines at the bottom corresponding to the transition from pure helium to the
helium--depleted core.   The profiles show very subtle changes
in the outer layers with evolution.}
\label{omvsrhb}
\end{figure}

The internal rotation profile remain nearly fixed
through the stage of core helium burning.  As shown in Figure~\ref{omvsrhb},
the rotation profile (with radius) shows the same general features at the
start and end of HB evolution.  Further evolution results in little additional angular velocity change with
radius.  This relatively small change in angular velocity profile on 
the EHB could allow efficient parameterization for studies of their seismic 
influence.

\section{Asteroseismology Probes}

Nonradial pulsations in sdB stars
provide a window into their interiors, and may test the hypothesis that
horizontal branch stars have rapidly rotating interiors.  Here we illustrate the process by which
we can detect rapid rotation of the stellar interior using currently observed
sdBV stars.  A future paper will describe our asteroseismic results in more
detail and apply the analysis to individual pulsators.

\begin{figure}
\centerline{\includegraphics[width=3.5in]{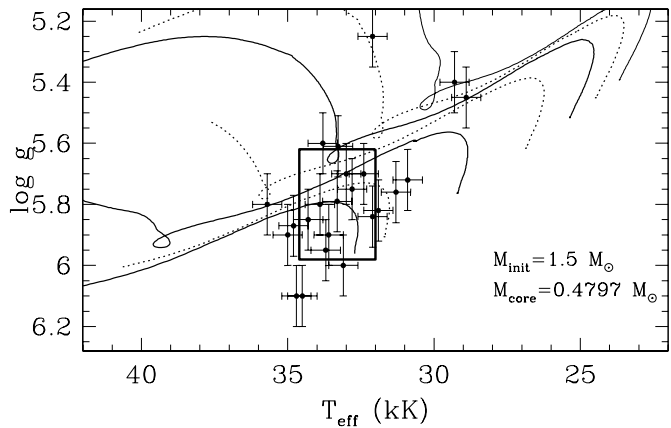}}
\caption{Evolutionary tracks of EHB models.  Models derived from the 
post-helium flash core
of a 1.5 $M_{\odot}$.   
Alternating tracks can be distinguished by
the different line types.  Envelope masses are 0.00001, 0.00004, 0.00032,
0.00122, 0.00232, 0.00322, 0.00422 $M_{\odot}$, running from lower left
to upper right.  Data from Kilkenny (2001).}  
\label{hrdsdbv}
\end{figure}

Figure~\ref{hrdsdbv}
shows an H--R diagram that includes the known sdB stars and sample
evolutionary tracks from our sets of models.  Given the convergence of
the tracks in this plane, it is clear that H--R diagram position alone is
insufficient to judge their evolutionary state, with stars that are close
to the ZAEHB mingling with stars that have depleted helium in their cores.
The lowest gravity EC~14026 stars fall sufficiently far from the ZAEHB that
they cannot be core helium--burners for core masses that are consistent with
single--star evolution.


Many of the high--gravity pulsating sdB stars show complex pulsations that
require a large number of periodicities.  Examples include
PG~1047+003 \inlinecite{Kilk1047} and PG~0014+067 \inlinecite{Brass0014}.  
Low $\ell$ nonradial modes have a limited
frequency distribution, meaning that some of the pulsators cannot be
understood without resorting to either high values of $\ell$
or to rapid rotation.

Nonradial oscillations are characterised by the
eigenfrequency $\sigma_{n \ell m}$ corresponding to a spheroidal mode with
quantum numbers $n$, $\ell$, and $m$.  The values $\ell$ and $m$ refer to the
spherical harmonic $Y^m_{\ell}$.  The sign of $m$ gives the direction of
propagation  of the surface running wave, and therefore there
are $2\ell+1$ possible values of $m$ for a given $\ell$.
For perfect spherical symmetry, the oscillation frequencies are independent
of $m$ and depend only on $n$ and $\ell$.  
But if rotation is present, the value of the oscillation frequency
will depend on $m$.  For
{\sl solid body} rotation the result of rotation is to split modes
of different values of $m$ by a constant frequency that is proportional to 
the rotation rate.  

If a star undergoes {\sl differential} rotation (i.e. $\Omega = \Omega
(r)$) then the splitting is given by
\begin{equation}
\sigma_{n \ell m} = \sigma_{n \ell 0} 
             - m \int_0^R{\Omega(r) K_{n \ell}(r) dr}
\end{equation}
where $K_{n \ell}(r)$
is the rotation kernel corresponding to a mode with $n$ and $\ell$. 
The observed splitting
for a mode with a given $n$ and $\ell$ represents an average of $\Omega(r)$ 
weighted by the rotation kernel $K_{n \ell}(r)$ \cite{KSG99}.  If the kernel function of a
mode is nonzero where the star has very rapid rotation, then the resulting
splitting can reveal the inner rapid rotation despite slow rotation at the 
surface layers.

\begin{figure}
\centerline{\includegraphics[width=4.2in]{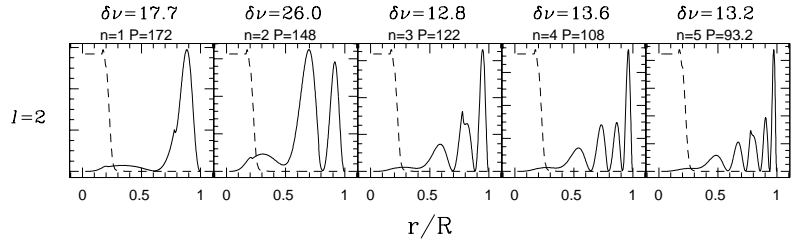}}
\caption{
Kernels for a sample of low $\ell$ nonradial modes in a Case B sdB model.
 The solid lines
represent the kernels, and the dashed line shows the (scaled) value of 
the angular rotation velocity as a function of fractional radius.  The rotational splitting for each mode is given above each panel
}  
\label{casebkern}
\end{figure}

Figure~\ref{casebkern} shows sample rotation kernels for an sdB model
representative of the pulsating EC 14026 stars.
The modes illustrated span the period range seen in the pulsating sdB stars.
Note that higher $n$ modes have more peaks, but that in all modes the
kernel has nonzero value close to the stellar core.  Also indicated in
Figure~\ref{casebkern} are lines representing $\log \Omega(r)$ for Case B
evolution of rotation in convection zones.  Note that even though the 
amplitude of $K_{n \ell}$ is small in
the inner regions, the large value of $\log \Omega(r)$ in the same region
suggests that the value of the rotational splitting will be dominated by
the core rotation.

For several $\ell=2$ modes in this model, the rotation kernel is significant
in the core.  This results from resonant mode trapping by the He/C+O
composition transition zone.  At this evolutionary stage, the model
shows some mixed--character modes for $\ell=2$ which can have significant
amplitude in the core \cite{CharpII}.

Figure 3 lists the computed splittings a Case B sdB model.  The surface
rotation velocity, if representative of the entire star, would produce
a splitting of less than $0.01 \mu{\rm Hz}$ for all modes (surface rotation
velocity of only 6 m/s), yet the rapid internal rotation results
in splittings ranging from 1 $\mu{\rm Hz}$ ($\ell=1$, $n=1$) to 
80 $\mu{\rm Hz}$ ($\ell=2$, $n=0$). 

In all cases, the rotational splitting is much larger than
expected from the surface rotation rate.  Also, the
splitting changes (sometimes dramatically) from
one mode to the next in a sequence of the same $\ell$ but different
$n$.  The
kernels (see Figure 3) sweeping
across the steep drop in $\Omega$ produce these fluctuations.  Mode trapping
by the composition transition zones produces some
strong core localization of the kernels for certain modes, and these also
show larger rotational splitting.

\section{Conclusions}

The sdB stars may retain rapidly rotating cores as a relic of their
evolution on the RGB.  Such rapid rotation could serve as a
reservoir of angular momentum which, when tapped, can produce
anomalously fast rotation at the surface on the HB.

Nonradial pulsations can reveal
the rapidly rotating cores via rotational splitting of nonradial
modes.  The rotational splitting will be much larger than expected
from the surface rotation velocity.  This may be happening in
PG~1605+072; the rotatonal splitting identified in that star by 
\inlinecite{Kaw98} is
about 3 times larger than the $v \sin i$ measured via spectroscopy
\cite{hebetal99}.  Also, the splittings seen in Feige 48
\cite{ReedetalF48} are significantly larger than expected from
the upper limits to $v \sin i$ by \inlinecite{hebetal99}.

The fact that the splittings can be quite large, coupled with the 
large differences in expected splittings from mode to mode, can make
observational identification of rotational splitting quite difficult.
Generally, one looks for
a sequence of equally--spaced triplets in a Fourier transform, and
identifies them as $\ell=1$ for example.  Even if all members 
of such triplets are not seen, finding several pairs of modes split
by the same amount would suggest rotational splitting.  For sdBV stars,
with significant mode--to--mode differences in splittings expected,
a series of rotationally split frequencies would be 
indistinguishable from modes of different $\ell$ and $n$. 
Without considering differential rotation, such a rich mode
spectrum would require appealing to other mechanisms.

Finally, we note that the periods seen in the PG1716+426 (``Betsy'') stars
are about the same periods that we expect for the rotation rates of 
the cores of sdB stars that evolve as single stars from the RGB.
Could this mean that the mechanism responsible for the pulsation
of these stars might be connected with rotation ($r$-modes or 
overstable convective modes)?

\begin{acknowledgements}
We appreciate support from the US National Science Foundation through Grant AST-0205983,
and to NASA for support through the Astrophysics Theory Program via grant NAG5-8352 to Iowa State University.
\end{acknowledgements}

\end{article}
\end{document}